\documentclass[aps,twocolumn,superscriptaddress]{revtex4-1}

\usepackage{breqn}
\usepackage{graphicx}
\usepackage{dcolumn}   
\usepackage{bm}        
\usepackage{amssymb}   
\usepackage{soul}
\usepackage{gensymb}
\usepackage{enumerate}
\usepackage{multirow}

\usepackage[breaklinks=true,colorlinks=true,linkcolor=blue,urlcolor=blue,citecolor=blue]{hyperref}

\usepackage{float}

\begin{document}

\title{Predictions of improved confinement in SPARC via energetic particle turbulence stabilization}

\author{A.~Di Siena} 
\affiliation{Max Planck Institute for Plasma Physics Garching 85748 Germany}
\author{P.~Rodriguez-Fernandez} 
\affiliation{MIT Plasma Science and Fusion Center Cambridge 02139 MA United States of America}
\author{N.~T.~Howard} 
\affiliation{MIT Plasma Science and Fusion Center Cambridge 02139 MA United States of America}
\author{A.~Ba\~n\'on~Navarro}
\affiliation{Max Planck Institute for Plasma Physics Garching 85748 Germany}
\author{R.~Bilato} 
\affiliation{Max Planck Institute for Plasma Physics Garching 85748 Germany}
\author{T.~G{\"o}rler} 
\affiliation{Max Planck Institute for Plasma Physics Garching 85748 Germany}
\author{E.~Poli} 
\affiliation{Max Planck Institute for Plasma Physics Garching 85748 Germany}
\author{G.~Merlo} 
\affiliation{The University of Texas at Austin Austin TX 78712 USA}
\author{J.~Wrigh} 
\affiliation{MIT Plasma Science and Fusion Center Cambridge 02139 MA United States of America}
\author{M.~Greenwald} 
\affiliation{MIT Plasma Science and Fusion Center Cambridge 02139 MA United States of America}
\author{F.~Jenko} 
\affiliation{Max Planck Institute for Plasma Physics Garching 85748 Germany}

\begin{abstract}

The recent progress in high-temperature superconductor technologies has led to the design and construction of SPARC, a compact tokamak device expected to reach plasma breakeven with up to $25$MW of external ion cyclotron resonant heating (ICRH) power. This manuscript presents local (flux-tube) and radially global gyrokinetic GENE (Jenko et al 2000 Phys. Plasmas {\bf 7} 1904) simulations for a reduced-field and current H-mode SPARC scenario showing that supra-thermal particles - generated via ICRH - strongly suppress ion-scale turbulent transport by triggering a fast ion-induced anomalous transport barrier (F-ATB). The trigger mechanism is identified as a wave-particle resonant interaction between the fast particle population and plasma micro-instabilities (Di Siena et al 2021 Phys. Rev. Lett. {\bf 125} 025002). By performing a series of global simulations employing different profiles for the thermal ions, we show that the fusion gain of this SPARC scenario could be substantially enhanced up to $\sim 80\%$ by exploiting this fast ion stabilizing mechanism. A study is also presented to further optimize the energetic particle profiles, thus possibly leading experimentally to an even more significant fusion gain. 

\end{abstract}


\maketitle


\section{Introduction}

Magnetic confinement research is steadily progressing towards the design and construction of reactor devices based on different theoretical and technological concepts. Among the most prominent examples, there are "low" magnetic field and large volume tokamak/stellarator (e.g., ITER and DEMO) \cite{Bigot_NF_2022,Siccino_FED_2022} or spherical tokamak (e.g., STEP) devices \cite{Wilson_BC_2020} which should operate at magnetic fields below 6T. However, major advancements in high-temperature superconductors (HTS) \cite{Whyte_JFE_2016,Cohn_JFE_1988} have led to the emergence of alternative reactor designs with high magnetic field and reduced plasma volume, which are particularly attractive from an economic point of view. The first breakeven-level machine of this kind - SPARC - is currently under construction in the USA \cite{Creely_JPP_2020}. SPARC is expected to reach plasma breakeven in a deuterium-tritium plasma heated by ion-cyclotron-resonant heating (ICRH) and lead to an overall plasma gain $Q > 2$. Different operational scenarios have been designed at SPARC to meet this outstanding goal. One of those is the so-called primary reference discharge (PRD) which is predicted to produce a fusion gain of $Q \approx 10$ in H-mode with full-field ($B_t = 12.2$T) and full-current ($I_p = 8.7$MA) with $^3$He as minority species and $11.1$MW of ICRH power. These predictions - initially based on scaling laws and simplified physical models - were recently corroborated via high-fidelity gyrokinetic simulations evolving plasma profiles and turbulence matching the volume integral of the injected sources and self-consistent exchange, radiation and alpha power predicting $Q \sim 8$ \cite{Rodriguez_Fernandez_NF_2022}.

In this manuscript, we focus on a different scenario characterized by a reduced magnetic field ($B_t = 8.5$), reduced current ($I_p = 6.1$MA) and hydrogen as minority species with an ICRH injected power of $P_{\rm ICRH} = 25$MW. This scenario is particularly interesting due to possible resonances between the  diamagnetic frequencies of the ICRH generated energetic particles and the bulk ion temperature gradient (ITG) driven micro-instabilities \cite{DiSiena_NF_2018,DiSiena_PoP_2019}. In particular, as discussed in Ref.~\cite{DiSiena_NF_2018,DiSiena_PoP_2019} fast particles can resonate with the underlying ITG micro-instabilities when their magnetic drift frequency is close to the ITG frequency. When this condition is fulfilled, a net energy transfer between the fast particle population and the ITG can occur, and - if the fast ion drive is properly optimized - ITG can be almost totally suppressed until possibly triggering an internal transport barrier \cite{DiSiena_PRL_2021,Di_Siena_PPCF_2022}. This is a novel type of transport barrier - so-called fast ion-induced anomalous transport barrier (F-ATB) - initially identified via global gyrokinetic simulations of an ASDEX Upgrade discharge, properly optimized to maximize fast ion effects on ion-scale plasma turbulence, showing features of improved plasma confinement \cite{DiSiena_PRL_2021}.

Detailed studies on this particular effect can be found in Ref.~\cite{DiSiena_PoP_2019}, where this mechanism is studied for different plasma compositions, plasma profiles \cite{Di_Siena_PPCF_2022} and magnetic geometries \cite{DiSiena_PRL_2020}. In the specific case of a hydrogen minority, the strongest ITG stabilization is typically observed when (i) the fast ion temperature is $40-80$keV, and (ii) their logarithmic temperature gradient largely overcomes the density gradient.

These conditions are typically met by energetic particles generated via ICRH, thus making SPARC - solely externally heated by ICRH - an ideal device to investigate such fast particle resonant effects. In particular, given the large ICRH power available at SPARC, and the medium-size plasma volume, a large fraction of fast particles with these desired conditions can also be generated at mid-radius. This largely broadens the operational regimes for these wave-particle effects and furthermore increases the experimental diagnostic coverage. It is worth mentioning here that fast ions with the optimal properties for suppressing ITG via resonant effects are created in a large fraction of the plasma volume for the reduced-field and current scenario, but only in a narrow layer close to the magnetic axis for the PRD scenario. This is due to the essentially on-axis ICRF heating applied for the PRD scenario. Similar mechanisms involving wave-particle interactions have also been derived for studying particle transport \cite{Angioni_NF_2012} and RSAE drive on thermal ions \cite{Nazikian_PRL_2006}.

Based on predictive TRANSP simulations with EPED-TGLF, the here considered off-axis ICRF heating scenario is predicted to reach a fusion gain of $Q \sim 0.9$. In the following, we will present dedicated local  and radially global gyrokinetic GENE \cite{Jenko_PoP2000,Goerler_JCP2011} simulations demonstrating that energetic particle effects - only partially captured by TGLF \cite{Doerk_NF_2017,Mantica_PPCF_2019,Reisner_NF_2020,Luda_NF_2021,DiSiena_NF_2022} when run with unusual high resolution \cite{Bilato_2022}) - strongly suppress turbulent transport in a narrow core region and ultimately trigger a transport barrier.

By performing a series of global gyrokinetic simulations with different temperature profiles for the thermal ions, this manuscript shows that the energetic particle turbulence suppression might potentially lead to a strong peaking of the thermal ion temperatures. As a main consequence of the increased on-axis temperature, the resulting fusion gain raises by $\sim 80\%$ up to $Q \sim 1.6$. Moreover, we suggest guidelines on possibly enhancing this fast particle stabilization to meet the SPARC mission goal of $Q \geq 2$.

This paper is organized as follows. The physical parameters, plasma profiles and geometry are discussed in Section \ref{sec1} where the reduced-field and current SPARC H-mode scenario is introduced. While section \ref{sec2a} briefly presents the gyrokinetic code GENE, the numerical setup and grid resolution employed for the simulations discussed throughout this paper are summarized in Section \ref{sec2b}. The role of energetic particles on this SPARC scenario is investigated with linear (local) and nonlinear (global and local) gyrokinetic simulations, respectively, in Section \ref{sec3a} and Section \ref{sec3b}, where a transport barrier is clearly observed in the simulations retaining the energetic particle species. In Section \ref{sec4} we perform a series of global nonlinear gyrokinetic simulations with different thermal ion (deuterium and tritium) temperature profiles keeping all other profiles fixed to the nominal ones, to identify the possible increase of the on-axis ion temperature due to the formation of the transport barrier. Due to the related computational burden, the simulations described in \ref{sec3} - \ref{sec4} are electrostatic. In Section \ref{sec5} the impact of electromagnetic fluctuations on the previous findings is assessed. In Section \ref{sec6} the optimal energetic particle parameters (density, temperature and logarithmic temperature gradients) to maximize their beneficial effect on ITGs is identified via linear (flux-tube) and nonlinear (global) simulations. Finally, conclusions are drawn in Section \ref{sec7}.

\section{Baseline H-mode off-axis SPARC scenario} \label{sec1}

The reference scenario analyzed throughout this manuscript has been designed to move the ICRF deposition layer to mid-radius $\rho_{tor} \approx 0.5$. This was done to generate an energetic particle population satisfying the optimal requirements for suppressing ITG turbulence via wave-particle resonant effects at the center of the plasma radial domain. To this end, a reduced magnetic field (compared to the PRD) with $B_t = 8.5$T and a constant ICRH antenna frequency of $\nu = 120$MHz leads to minority effective temperatures of $T_h >40$keV near $\rho_{tor}=0.5$ with $P_{\rm ICRH} = 25$MW of input power according to TORIC/SSFPQL \cite{Bilato_NF_2011} simulations. A reduced toroidal current (compared to the PRD) of $I_p = 6.1$MA is here employed leading to the same value of $q_{95} = 3.4$ predicted for the PRD.

The considered plasmas contain deuterium, tritium, electrons (here modeled with realistic ion-to-electron mass ratio) and a fast ion hydrogen minority heated via ICRH. The hydrogen concentration is fixed to $n_h / n_e = 0.055$ and impurities are neglected in our modelling to reduce the computational resources required for the GENE simulations. The plasma profiles are computed via TRANSP-TGLF simulations, with a fixed pedestal top temperature as calculated by EPED with the same assumption for the Greenwald fraction as in the PRD scenario. All the predictive modeling with TRANSP was done in a similar way as presented in Ref.~\cite{Rodriguez_Fernandez_JPP_2020} for the PRD scenario.

\begin{figure*}
\begin{center}
\includegraphics[scale=0.35]{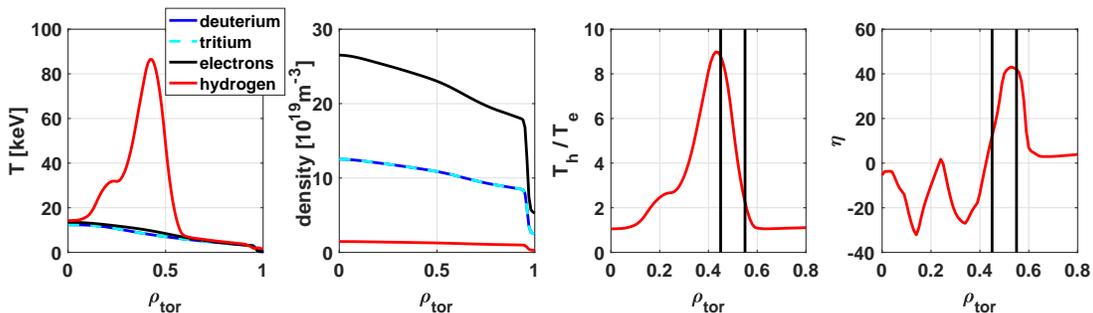}
\par\end{center}
\caption{(a) Temperature profiles, (b) density profiles, (d) ratio between hydrogen and electron temperatures $T_h / T_e$ and ratio of logarithmic temperature ($\omega_{T_h} = a/L_{T_h}$) and density ($\omega_{n_h} = a/L_{n_h}$ with a minor radius) gradients of the hydrogen minority $\eta = \omega_{T_h} / \omega_{n_h}$. The vertical black lines in (c) and (d) delimit the radial domain where the fast ion (hydrogen minority) species reach the optimal parameters for suppressing ITGs via wave-particle resonant effects.}
\label{fig:fig1}
\end{figure*}

The temperature and density profiles of the plasma species considered in the GENE simulations are shown in Fig.~\ref{fig:fig1}. The ratio between the electron kinetic and magnetic pressure ($\beta_e$) measures $\beta_e \sim 0.009$ at mid radius.
Furthermore, the hydrogen to electron temperature ratio and the ratio between the hydrogen logarithmic temperature and density gradients ($\eta = \omega_{T,h} / \omega_{n,h}$, with $\omega_{n,h} = -(a / n_h)$ d$n_h /$d$ \rho_{tor}$ - and same definition for the temperature - with $a$ minor radius) have been added in Fig.~\ref{fig:fig1} to better assess the radial domain where wave-particle resonant effects are expected to be most effective. According to previous theoretical and numerical analyses on different plasma conditions, these optimal fast ion parameters to maximize the resonant effects for hydrogen minority are $T_h / T_e = [3 - 6]$ and large $\eta$ \cite{DiSiena_PoP_2019}. These requirements are reached in the radial domain $\rho_{tor} = [0.45 - 0.55]$ (see Fig.~\ref{fig:fig1}). For $\rho_{tor}> 0.55$, the hydrogen temperature reduces to the electron one, and the wave-particle resonant interaction is expected to turn from stabilizing to destabilizing ITGs. For $\rho_{tor}<0.4$, the optimal temperature ratio is also achieved, but the negative logarithmic fast ion temperature gradient changes the sign of the fast particle drive term in phase-space. As shown in Ref.~\cite{DiSiena_PoP_2019}, this creates a radial layer where the wave-particle interaction enhances the ITG drive. Therefore, localized layers, where ITGs are destabilized, are expected with a strong stabilization within these layers, thus possibly triggering an internal transport barrier as observed in Ref.~\cite{DiSiena_PRL_2021,Di_Siena_PPCF_2022}.

\section{Simulation details}

\subsection{Code description} \label{sec2a}

The numerical simulations presented in this work have been performed with the Eulerian gyrokinetic code GENE \cite{Jenko_PoP2000,Goerler_JCP2011}. GENE solves the Vlasov-Maxwell equations for the perturbed part of the distribution functions of the different plasma species considered, assuming a constant background. While a typical choice is given by a Maxwellian distribution (as employed throughout this paper), more sophisticated backgrounds are also supported in the code \cite{DiSiena_PoP_2018}. Realistic collisional operators (e.g., linearized Landau operator \cite{Merz_phd_thesis}, Sugama collisional model \cite{Crandall_CPC_2020} and exact Fokker-Planck operator \cite{Qingjiang_PRE_2021}), electromagnetic fluctuations and interfaces to MHD equilibrium codes, e.g., CHEASE \cite{Lutjens_CPC_1996}, are also available. 

The underlying equations are discretized on the field-aligned coordinate grid $\left(x,y,z\right)$ in space and $\left(v_\shortparallel, \mu\right)$ in velocity with $x = \rho_{tor} = \sqrt{\Phi_{tor}/\Phi_{{\rm edge}}}$ the radial, $y = x_0\left(q\chi -\varphi \right)/q(x_0)$ the bi-normal and $z$ the straight-field line directions. Here, $\Phi_{tor}$ represents the toroidal flux, $\Phi_{{\rm edge}}$ is its value at the last closed flux surface, $B_0$ the on-axis magnetic field, $x_0$ the center of the radial domain, $\chi$ the straight-field-line poloidal angle, $\varphi$ the toroidal angle and $q$ the safety factor.

Moreover, $v_\shortparallel$ and $\mu$ represent, respectively, the velocity component parallel to the background magnetic field and the magnetic moment. Equidistant grids are used for the field-line z and $v_\shortparallel$ directions and Gauss-Laguerre integration points for the magnetic moment. The bi-normal direction ($y$) is represented in Fourier space. While an equidistant grid is also used for the radial direction ($x$) in the radially global version of the code, a Fourier decomposition is employed when only a single flux-tube (local) is considered. To reduce the computational cost GENE supports block-structured velocity grids, where the box size in $(v_\shortparallel, \mu)$ is properly adjusted along the radial direction to minimize the resolution requirements \cite{Jarema_CPC_2016}.

While periodic boundary conditions are applied in flux-tube simulations, radially global simulations require Dirichlet boundary conditions, which are enforced with radial buffer regions to damp fluctuations close to the boundaries. Another difference between flux-tube and global simulations is that the latter require Krook-type operators to maintain the averaged profiles of each species fixed to the initial ones. Further details on the numerical schemes and implementation in GENE can be found in Ref.~\cite{Goerler_JCP2011}.

\subsection{Numerical setup and resolution} \label{sec2b}

In the present work, the gyrokinetic code GENE has been used both in the radially global and flux-tube (local) limit. Here, we summarize the numerical parameters and grid resolutions employed for our numerical simulations. All simulations are performed retaining kinetic electrons with realistic ion-to-electron mass ratio, collisions modeled with a linearized Landau operator with energy and momentum conserving terms and the equilibrium is reconstructed by TRACER-EFIT \cite{Xanthopoulos_PoP_2009}. Impurities are not retained in the GENE simulations due to the otherwise prohibitive computational cost to include an additional plasma species.

The GENE global simulations are performed in the radial domain $\rho_{tor} = [0.25-0.8]$. To damp fluctuations and enforce Dirichlet boundary conditions, buffer regions are applied covering $10\%$ of the GENE radial domain. In these regions, artificial Krook operators are applied with a relaxation rate of $\gamma_l = 1.0 c_s/a$ for the inner buffer region and $\gamma_u = 7.0 c_s/a$ for the outer one. Here, $c_s = (T_e / m_i)^{1/2}$ is the sound speed, with $m_i$ the bulk ion mass in proton units and $T_e$ the electron temperature at the center of the radial domain. Fine electron scale turbulence due to electron temperature gradient (ETG) modes is artificially damped via numerical fourth order hyperdiffusion. Finally, to reduce the resolution requirements in the velocity space, a radially dependent block-structured grid is employed in the global GENE simulations. It is divided into five different blocks, properly built to resolve correctly the dynamics of all the plasma species considered with the selected grid resolutions.

The grid resolutions and box sizes employed for the global and flux-tube (local) simulations are summarized in table \ref{table:tab1}. The discretized toroidal mode number is given by $n = n_{0,min} \cdot j$ with $j$ being integer-valued in the range $j = [0,1,2, ..., n_{ky0}]$. The simulations in Section \ref{sec3} - \ref{sec4} are performed in the electrostatic limit while the ones in section \ref{sec6} are electromagnetic.
\begin{table}
\centering
\begin{tabular}{ ||c| c| c|| }
 \hline
 GENE setup & flux-tube (local) & global\\
 \hline
 $n_x \times n_{ky0} \times n_z$   & $195 \times 64 \times 24$    & $512 \times 64 \times 48$\\
 $n_{v_\shortparallel} \times n_{\mu}$   & $32 \times 24$    &$48 \times 30$ \\
  $n_{0,min}$    &  4  &6   \\
  $l_{v_{min}}, l_{v_{max}}$    &  -3,  3  &  -3.5, 3.5\\
  $l_{w_{min}}, l_{w_{max}}$    &  0, 9 &  0, 11.5 \\
 \hline
\end{tabular}
\caption{GENE grid resolution and box sizes. Here, $n_x$, $n_{ky0}$ and $n_z$ denote, respectively, the numerical resolution used along the radial, bi-normal and field-aligned directions; $n_{v_\shortparallel}$ and $n_{\mu}$ the resolution used in velocity space for the velocity component parallel to the background magnetic field and the magnetic moment; $l_{v_{min}}, l_{v_{max}}$ and $l_{w_{min}}, l_{w_{max}}$ denote, respectively, the extension (minimum/maximum value) of the simulation box in the $v_\shortparallel$ and $\mu$ directions in units of the thermal velocity of each species. While the thermal velocity is computed at the considered location for flux-tube runs, it is taken at the center of the radial grid in the global simulations.}
\label{table:tab1}
\end{table}

\section{Wave-particle resonant effects in SPARC} \label{sec3}

\subsection{Local linear simulation results} \label{sec3a}

\begin{figure}
\begin{center}
\includegraphics[scale=0.3]{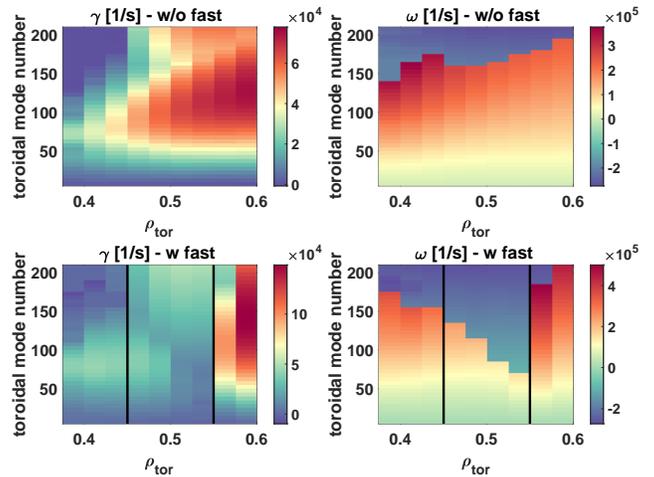}
\par\end{center}
\caption{Contour plots of the most unstable linear growth rate and frequency (in units of $1/s$) for the toroidal mode numbers $n = [1-200]$ at different radial locations (a) neglecting and (b) retaining the energetic particle species in the GENE linear simulations. The vertical black lines delimit the region where the fast ion temperature and its logarithmic gradient reach the optimal values (see Fig.~\ref{fig:fig1}) for efficiently suppressing ITG micro-instabilities.}
\label{fig:fig2}
\end{figure}

We begin our analyses by investigating the impact of the ICRH hydrogen minority species (energetic particles) on the linear micro-instabilities. Given the primary role played by ion-scale plasma turbulence in SPARC \cite{Rodriguez_Fernandez_JPP_2020,Howard_PoP_2021}, we restrict our analyses to the toroidal mode numbers $n = [5 - 200]$, where ITG is the dominant instability. In particular, a series of flux-tube simulations at different radial locations are performed in the electrostatic limit, retaining and neglecting the hydrogen minority in the modelling. The plasma profiles are those discussed in Sec.~\ref{sec2b}. Plasma quasi-neutrality is ensured in the runs without energetic particles on the electron density. The results are shown in Fig.~\ref{fig:fig2}, where we compare the linear growth rates obtained in the simulations without (Fig.~\ref{fig:fig1}a) and with (Fig.~\ref{fig:fig1}b) energetic particles.
In the absence of fast ions (hydrogen minority), smooth growth rate variations are observed along the radial direction, showing a maximum at approximately $n \approx 100$. 

When the energetic particles are included in the simulations, we observe a sharp drop of the ITG growth rates in $0.45 <\rho_{tor} < 0.55$. This is the radial region where the hydrogen temperature and its logarithmic gradient reach the optimal conditions for maximizing the beneficial effect of the wave-particle resonant interaction on ion-scale instabilities \cite{DiSiena_NF_2018,DiSiena_PoP_2019}. Outside this radial region towards the edge, the wave-particle mechanism turns from stabilizing to destabilizing (due to the almost thermal fast ion temperature), thus leading to an ITG destabilization (at $\rho_{tor} \sim 0.55$).
\begin{figure}
\begin{center}
\includegraphics[scale=0.37]{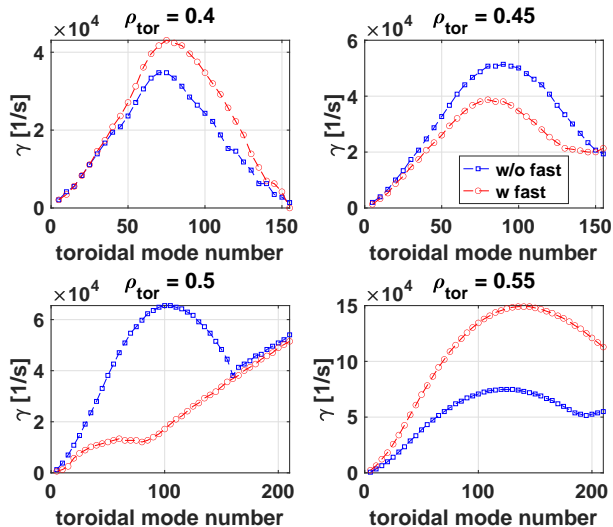}
\par\end{center}
\caption{Slices of Fig.~\ref{fig:fig2} at different radial locations where the most unstable linear growth rates (in unit of $1/s$) are obtained in the simulations retaining (red circles) and neglecting (blue square) the fast particle species are compared.}
\label{fig:fig3}
\end{figure}
This is shown more clearly in Fig.~\ref{fig:fig3}, where the ITG linear spectra are illustrated at different radial location, namely $\rho_{tor} = [0.4,0.45,0.5,0.55]$. 

In particular, we notice a mild fast ion destabilization at $\rho_{tor} = 0.4$. While this is a location where $T_f / T_e \approx 8$ is large enough to move the resonant layer in phase-space in the optimal regions, the logarithmic fast ion temperature gradient is weakly negative. As discussed in Ref.~\cite{DiSiena_PoP_2019}, the negative logarithmic temperature gradient leads to a change in the direction of the net energy transfer between fast particles and the wave and thus a linear ITG destabilization. At $\rho_{tor} = 0.45$, the nearly zero logarithmic fast ion temperature gradient makes the energy transfer particularly ineffective, resulting only in a weak ITG stabilization. The optimal combination of $T_f / T_e = [3 - 6]$ and logarithmic fast ion temperature gradient $\omega_{T,h}$ is reached at $\rho_{tor} = 0.5$. Here, the fast ion magnetic drift frequency matches the ITG frequency for a broad range of mode numbers, and the large logarithmic gradient strongly enhances the energy exchange from the instability to the fast ions. This leads to almost a full ITG suppression. Interestingly, as the ITG growth rate is reduced, trapped electron modes (TEM) become the dominant instability at $n > 100$. The wave-particle resonant interaction does not impact effectively electron driven instabilities and no differences in the linear growth rates are observed at $\rho_{tor} = 0.5$ for $n > 150$.

At $\rho_{tor} = 0.55$, the fast ion temperature is almost thermal, and the resonant layers move in the destabilizing phase-space region. At this location, the large (positive) fast ion logarithmic temperature gradient strongly enhances the energy transfer from the particles to the ITGs, leading to a large linear destabilization.


These findings are well consistent with theoretical predictions \cite{DiSiena_NF_2018}, previous gyrokinetic simulations \cite{DiSiena_PoP_2019} and experimental results \cite{DiSiena_PRL_2021}. We refer the reader to Refs.~\cite{DiSiena_PoP_2019} for a detailed description of the physical mechanism responsible for the interplay between stabilizing/destabilizing effects on the underlying ITB micro-instabilities by energetic particles.

\subsection{Global nonlinear simulation results} \label{sec3b}

Following the linear results discussed above, the impact of the wave-particle resonant interaction on this SPARC H-mode scenario is further addressed by performing global nonlinear turbulence simulations. To isolate the effect of the hydrogen minority on the thermal species fluxes, we retain and neglect the fast particles, similarly, as done in the previous section. We enforce plasma quasi-neutrality by modifying the electron density profile in the simulations without fast particles. The numerical setup and grid resolutions are the same as the one discussed in Sec.~\ref{sec3a}. 

\begin{figure*}
\begin{center}
\includegraphics[scale=0.45]{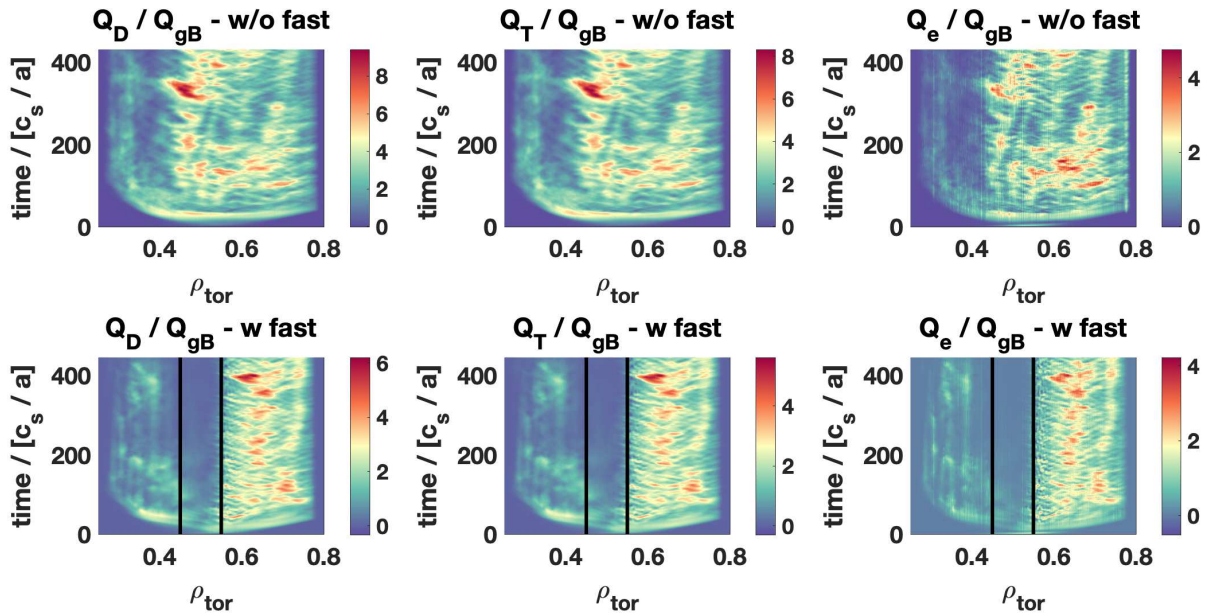}
\par\end{center}
\caption{Time evolution of the radial heat flux profile in gyro-Bohm units for (a) deuterium, (b) tritium and (c) electrons retaining (botton figures) and neglecting (top figures) the fast particle species. The vertical black lines delimit the region where the fast ion temperature and its logarithmic gradient reach the optimal values (see Fig.~\ref{fig:fig1}) for efficiently suppressing ITG micro-instabilities.}
\label{fig:fig4}
\end{figure*}

The time evolution of the radial profile of the thermal species heat fluxes are shown in Fig.~\ref{fig:fig4} for the simulations without (top plots) and with fast particles (bottom plots).
In the absence of the energetic particles (no hydrogen minority), turbulence avalanches propagate throughout the radial domain $\rho_{tor} = [0.5 - 0.75]$ for both thermal ions and electrons. These elongated structures break when fast particles are consistently included in the GENE simulations in $\rho_{tor} = [0.45 - 0.55]$. This is the radial region where the energetic particle temperature and its logarithmic gradient reach the optimal values for effectively suppressing ITG micro-instabilities (as shown in Sec.~\ref{sec3a}), thus leading to a significant turbulence suppression. Outside this region - specifically, at $\rho_{tor} \sim 0.55$ - the resonant interaction turns from stabilizing to destabilizing in a localized layer, delimiting the outer boundary of the transport barrier. 
\begin{figure}
\begin{center}
\includegraphics[scale=0.45]{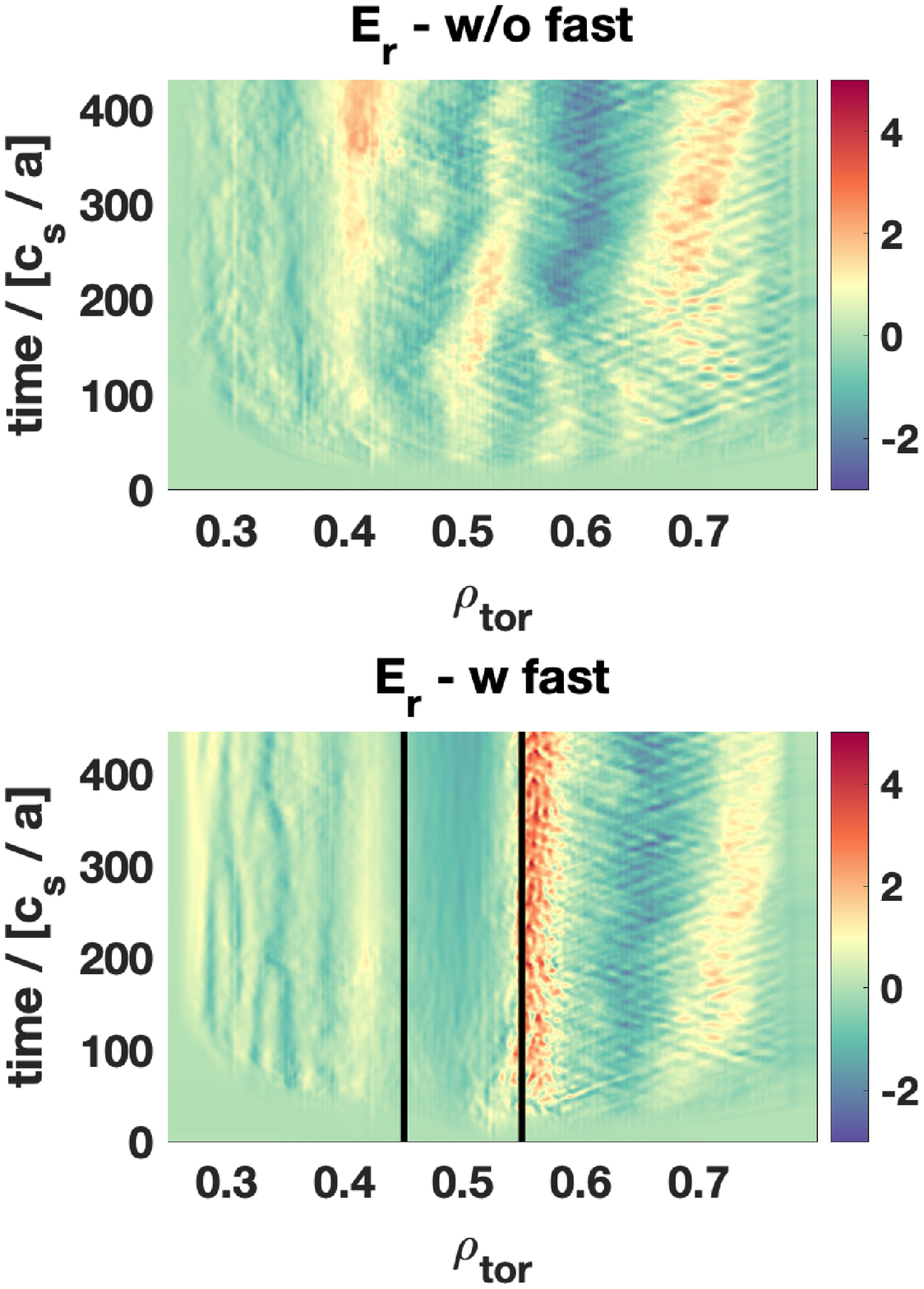}
\par\end{center}
\caption{Time evolution of the radial profile of the flux-surface averaged radial electric field $E_r = -\partial_{\rho_{tor}} \phi_1(n = 0)$ obtained neglecting (a) or retaining (b) the fast ion species. The vertical black lines delimit the region where the fast ion temperature and its logarithmic gradient reach the optimal values (see Fig.~\ref{fig:fig1}) for efficiently suppressing ITG micro-instabilities.}
\label{fig:fig5}
\end{figure}
At this position, the destabilizing effect that supra-thermal particles have on ion-scale micro-turbulence leads to the excitation of narrow zonal flow layers that help in triggering the transport barrier, as discussed in Ref.~\cite{DiSiena_PRL_2021}. This is shown in Fig.~\ref{fig:fig5}, where the spatial and temporal evolution of the flux-surface averaged radial electric field $E_r$ is illustrated for the GENE simulation retaining the energetic particles. 
%
%
These findings are consistent with Ref.~\cite{DiSiena_PRL_2021}, where this type of transport barrier - called fast ion-induced anomalous transport barrier (F-ATB) - was identified via global gyrokinetic simulations on a properly optimized experimental discharge at ASDEX Upgrade.

\begin{figure*}
\begin{center}
\includegraphics[scale=0.40]{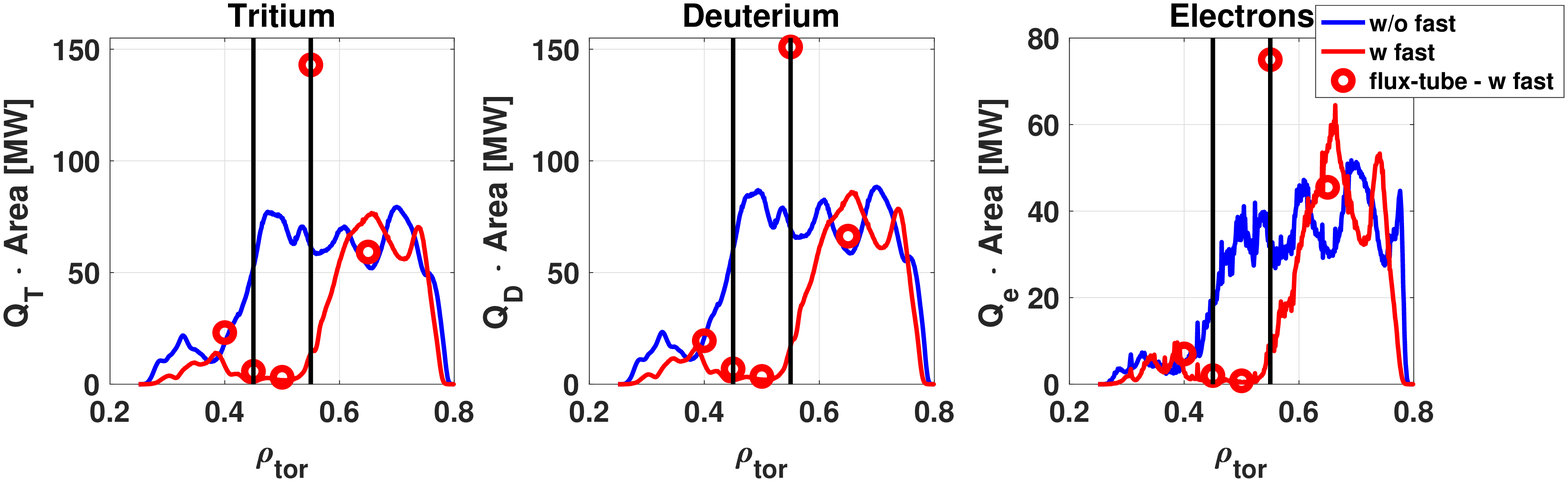}
\par\end{center}
\caption{Radial turbulent heat flux profiles obtained in the global GENE simulations for (a) tritium, (b) deuterium and (c) electrons averaged over the time domain $t[c_s/a] = [250–400]$ neglecting (blue) and retaining (red) the fast particle species. The red circles denote the turbulent fluxes obtained in the flux-tube simulations at different radial locations. The vertical black lines delimit the region where the fast ion temperature and its logarithmic gradient reach the optimal values (see Fig.~\ref{fig:fig1}) for efficiently suppressing ITG micro-instabilities.}
\label{fig:fig6}
\end{figure*}

For a quantitative comparison of the previous results, we add in Fig.~\ref{fig:fig6}, the time-averaged turbulent fluxes (for the thermal species) and flux-surface averaged radial electric field. In addition, Fig.~\ref{fig:fig6} contains the turbulent fluxes obtained via GENE flux-tube simulations at different radial locations by retaining and neglecting the energetic particles.
Comparing the global and the local results (blue crosses and red squares), we observe a good agreement for $0.4<\rho_{tor}\leq 0.5$, with differences of $\sim 15\%$ (comparable with the uncertainty of the average flux calculations). At $\rho_{tor} = 0.55$, the flux-tube simulation significantly over-predicts the turbulent fluxes by more than one order of magnitude. As mentioned previously, this is the location where the wave-particle resonant interaction leads to a linear ITG destabilization. However, while the increased ITG drive is observed together with a corresponding increase in the zonal flow levels in the global simulation, this is not captured correctly in the local framework. These results show that flux-tube (local) simulations fail to reproduce the global results at the proximity of the outer boundary of the transport barrier. More detailed analyses are required to better identify the limitations of the local approximation in modelling fast particle effects on turbulent transport. This will be addressed in a separate publication.

\section{Ion temperature peaking due to fast particle stabilization} \label{sec4}

Based on the linear (local) and nonlinear (global) GENE simulations performed in the previous sections, a fast ion-induced anomalous transport barrier might be triggered in this SPARC H-mode scenario. In particular, we observe (i) a significant stabilization of the linear ITG growth rates, (ii) a large drop of all turbulent fluxes and (iii) enhanced zonal flow activity in the region where the hydrogen minority is heated with ICRH. These findings can be explained in terms of fast particle ITG stabilization via the wave-particle resonant interaction identified in Ref.~\cite{DiSiena_PRL_2021}. Despite these standalone results (with TGLF-predicted gradients) show clearly that fast particles can be highly beneficial for this SPARC H-mode scenario, it is still unclear how their stabilizing effect affects the thermal ion temperature profiles and hence plasma performance. This is extremely important for fusion reactors, where the key goal is maximizing fusion energy output.

This is partially addressed in this Section where we perform a series of global nonlinear GENE turbulence simulations with different temperature profiles for the thermal ions (kept the same for deuterium and tritium) but keeping all the other profiles fixed to the ones discussed in Sec.~\ref{sec3}. Our aim is to identify the on-axis peaking of the ion temperature profile which could be achieved thanks to the fast ion turbulence stabilization observed in Sec.~\ref{sec3}. The reference heat fluxes taken for this analyses are the ones obtained in the absence of supra-thermal particles in Sec.~\ref{sec3b}. More specifically, we consider the plasma profiles computed by TRANSP-TGLF - which do not capture fast ion effects on turbulence correctly - as the ones matching the volume integral of the injected power when impurities and electromagnetic effects are included in the modelling. Therefore, we design different ion temperature profiles (for deuterium and tritium) and identify the one with turbulent fluxes comparable with the reference scenario (the case without fast particles studied in Sec.~\ref{sec3b}) as the enhanced profile. Modifications on the magnetic geometry due to the increase in the plasma pressure and its gradient are not considered in these analyses. 

\begin{figure}
\begin{center}
\includegraphics[scale=0.33]{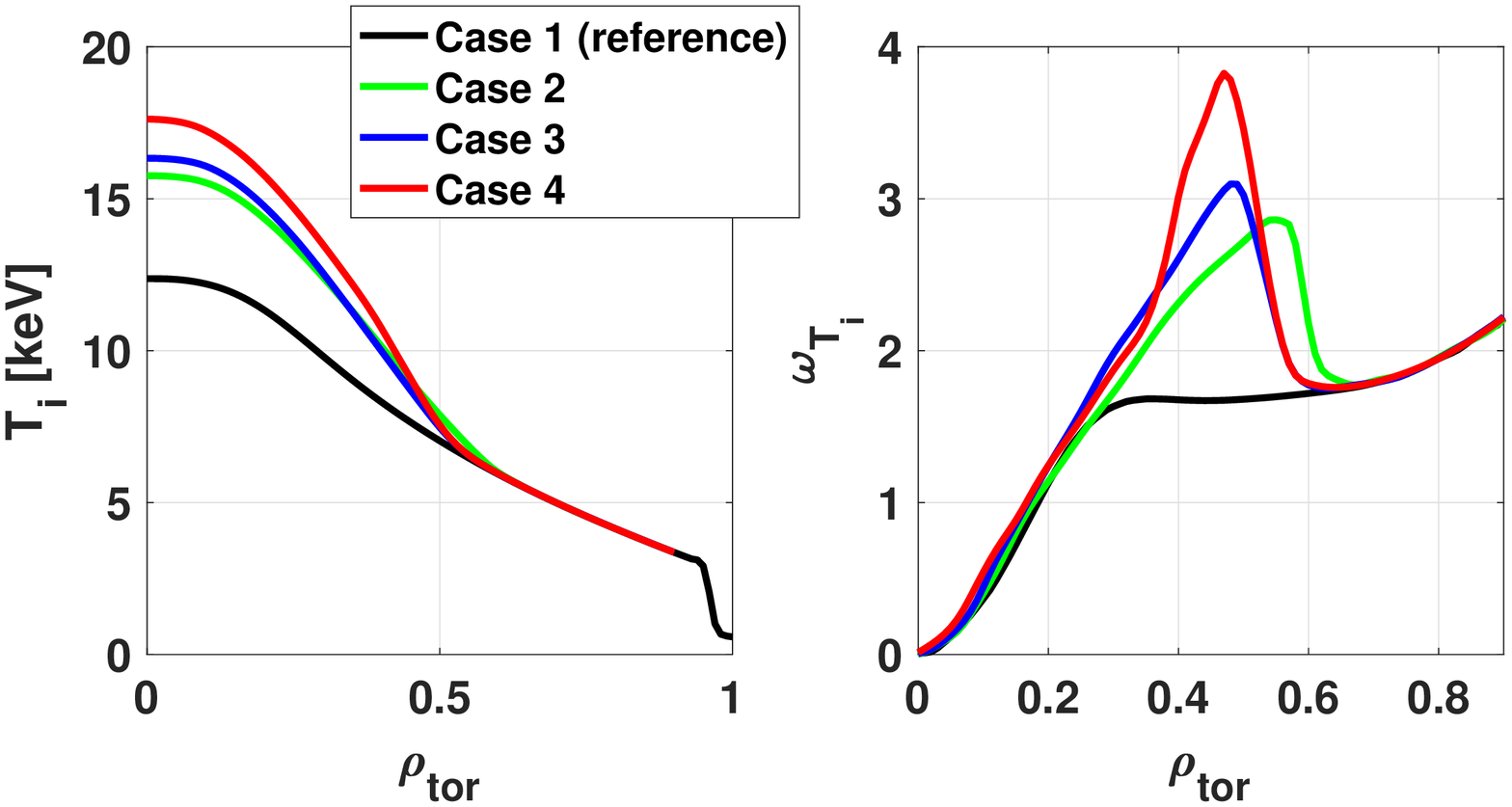}
\par\end{center}
\caption{Radial profiles of the deuterium and tritium (a) temperature and (b) logarithmic temperature gradients. The black line represents the reference profile computed by TRANSP (Case 1).}
\label{fig:fig7}
\end{figure}

\begin{figure*}
\begin{center}
\includegraphics[scale=0.40]{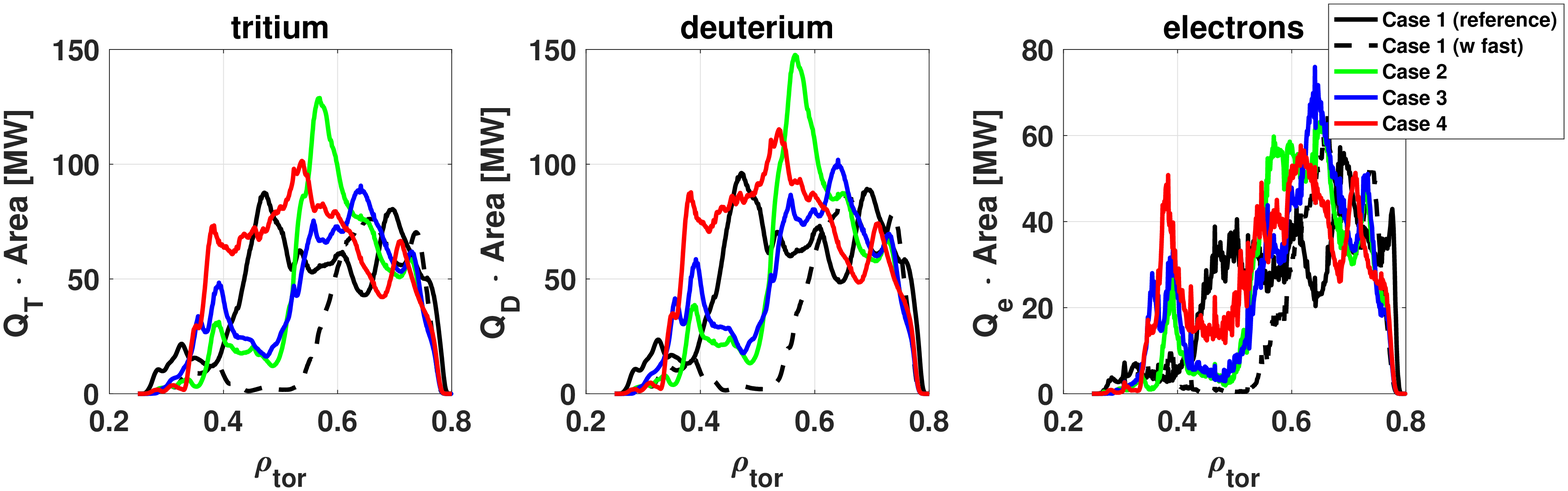}
\par\end{center}
\caption{Radial turbulent heat flux profiles obtained in the global GENE simulations for (a) tritium, (b) deuterium and (c) electrons for the four different thermal ion profiles of Fig.~\ref{fig:fig7} retaining fast particles. The heat fluxes are averaged over the time domain $t[c_s/a] = [250–400]$.}
\label{fig:fig8}
\end{figure*}

The thermal ion profiles employed for this study are shown in Fig.~\ref{fig:fig7} together with their logarithmic temperature gradient. 
For each of these different cases, we perform global electrostatic GENE simulations retaining the supra-thermal particle species (hydrogen minority). The results are summarized in Fig.~\ref{fig:fig8}, where the time-averaged turbulent fluxes are illustrated and compared with the reference case (Case 1) without fast particles. 

A first sticking observation when looking at Fig.~\ref{fig:fig8}, is the transport barrier strength. In particular, the sharp drop in the turbulent fluxes between $0.45<\rho_{tor}<0.55$ persists for the profiles labelled Case 2 and Case 3 despite the large increase in the thermal ion logarithmic temperature gradient (see Fig.~\ref{fig:fig7}). To reach turbulent levels comparable with the reference case without fast particles (Case 1), the logarithmic temperature gradient of the thermal ions is increased by more than a factor of two, resulting in a $\sim 40\%$ increased of the on-axis ion temperature. This corresponds to the case labelled Case 4 where fast particles are retained in the GENE simulations. These results are particularly promising for SPARC aiming to achieve $Q > 2$. 

More specifically, for this specific H-mode SPARC scenario with off-axis ICRF heating, the fusion gain computed by EPED-TGLF with the nominal plasma profiles is $Q \sim 0.9$. However, when taking the thermal ion profiles (Case 4) matching qualitatively the turbulent fluxes obtained in the GENE global simulations without fast particles, the fusion gain increase by $\sim 80\%$ reaching $Q \sim 1.6$ as computed on the profiles labelled Case 4 in Fig.~\ref{fig:fig7}.

It is worth mentioning that a quantitative assessment of the increase in the fusion gain due to the wave-particle resonant stabilization of ITG turbulence in this SPARC scenario would require flux-matching simulations, where the plasma profiles are allowed to self-consistently evolve in the gyrokinetic simulations due to the combined effect of energy sources (external heating, alpha heating, exchange and radiation) and plasma turbulence. This is essential to capture also modifications of the plasma heating due to the change in the plasma profiles. In particular, the increased on-axis thermal ion temperature leads - not only - to a corresponding improvement of the fusion output but it also affects the plasma heating (e.g., alpha particles heating). Additionally, changes on the electron temperature and density profiles - not considered here - should also be included in the modelling, which are required to achieve a better agreement for the electron heat flux between the reference case (Case 1) and Case 4 (see Fig.~\ref{fig:fig8}). However, flux-matching global simulations with high-fidelity (gyrokinetic) codes to the transport time scale is a major challenge given the prohibitive computational cost of such runs. Therefore, only a limited number of such simulations have been performed so far and mainly with reduced plasma setups  \cite{Dif-Pradalier_PRL_2015,Strugarek_PPCF_2013}. 

A large effort has been spent in recent years to overcome these limitations of gyrokinetic codes. One of the most promising approaches consists of coupling the gyrokinetic code to a transport solver and exploiting the time scale separation that exists between microscopic and macroscopic physics. Such coupling allows the gyrokinetic codes to run only over a few microscopic time steps to evaluate the turbulence fluxes for given plasma profiles, and afterwards, the transport solver evolves these plasma profiles to the next macroscopic time step. Substantial speedups have been reported in the literature with such code coupling, enabling profile predictions with gyrokinetic codes. This (or similar approaches) is done both with flux-tube \cite{Barnes_PoP_2010,Rodriguez_Fernandez_NF_2022} and radially global gyrokinetic codes \cite{DiSiena_NF_2022}.

In the near future, flux-matching simulations will be performed on this specific scenario to refine the prediction of a $\sim 80\%$ increase in the fusion gain as discussed in this section.

\section{Impact of electromagnetic fluctuations} \label{sec5}

The numerical simulations discussed in the previous sections were all performed in the electrostatic limit. However, supra-thermal particles are known to impact ion-scale turbulent transport via electromagnetic effects (e.g., nonlinear coupling with fast ion driven electromagnetic modes, mode conversion to electromagnetic instabilities, and modifications on the magnetic geometry) \cite{Bourdelle_NF_2005,MRomanelli_PPCF2010,Citrin_PRL_2013,DiSiena_NF_2019}. Moreover, energetic particles could excite electromagnetic modes that might be detrimental for plasma confinement \cite{Citrin_PPCF_2014, DiSiena_JPP_2021, Biancalani_PPCF_2021, Ishizawa_NF_2021}.

In this section, we further extend our numerical simulations by comparing the electrostatic and the electromagnetic turbulent fluxes obtained in global gyrokinetic simulations, thus investigating whether electromagnetic fast ion effects might play a relevant role in this SPARC H-mode scenario. 

When retaining electromagnetic effects self-consistently in the GENE simulations, the computational cost of each global nonlinear run strongly increases. Therefore, to reduce the otherwise prohibitive computational cost, we restrict this analysis only to the Case 3 described in Sec.~\ref{sec4}. The results of this Section do not change qualitatively if a different thermal profile - among the ones introduced in Sec.~\ref{sec4} - is selected. 

The electron kinetic-to-magnetic pressure ratio $\beta_e = 0.96\%$ (at the center of the GENE radial domain $\rho_{tor} = 0.525$) is consistently computed on the electron pressure profile (the latter kept the same for all the cases studied in this manuscript). The thermal ion beta profiles are self-consistently adjusted for each cases studied in this manuscript.

Moreover, modifications on the magnetic geometry due to the fast particle pressure are included in our modelling. 
\begin{figure*}
\begin{center}
\includegraphics[scale=0.40]{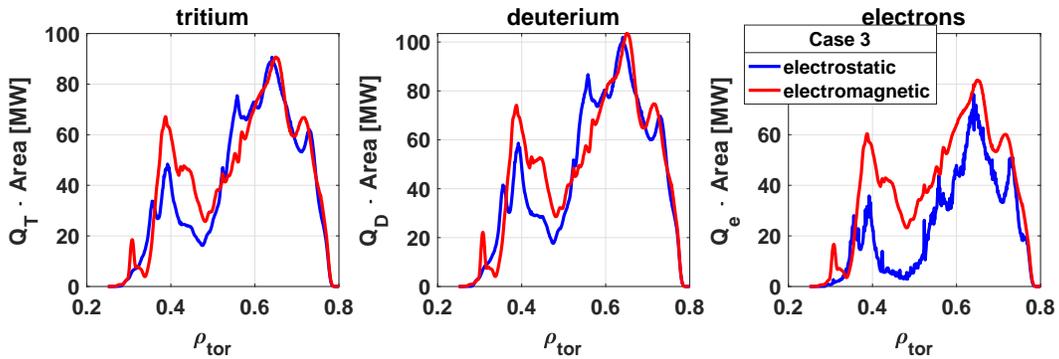}
\par\end{center}
\caption{Radial turbulent heat flux profiles obtained in the electrostatic (blue) and electromagnetic (red) global GENE simulations for (a) tritium, (b) deuterium and (c) electrons retaining fast particles for the thermal ion profile labelled Case 3 in Fig.~\ref{fig:fig7}. The heat fluxes are averaged over the time domain $t[c_s/a] = [250–400]$.}
\label{fig:fig9}
\end{figure*}
The results are shown in Fig.~\ref{fig:fig9}, where the time-averaged turbulent fluxes obtained from the electromagnetic simulations are illustrated and compared with the electrostatic results (Fig.~\ref{fig:fig8}.). The inclusion of electromagnetic fluctuations in the global GENE simulations does not lead to large modifications of the turbulent fluxes for the thermal ions (both deuterium and tritium). This is likely to be caused by the large magnetic field and relatively low electron kinetic-to-magnetic pressure ratio in the radial region where the fast particle population is created. This reduces the drive of electromagnetic fast particle modes and weakens their nonlinear coupling with ITG turbulence, often considered as the main electromagnetic fast ion stabilizing effect on ITG turbulence \cite{DiSiena_NF_2019}. However, we notice a corresponding increase of the electron turbulent flux in the electromagnetic simulation, which is consistent with the numerical analyses reported in Ref.~\cite{DiSiena_JPP_2021}.

The overall weak impact of electromagnetic fluctuations on ion-scale plasma turbulence on SPARC is consistent with previous simulations of the primary reference discharge (PRD) \cite{Rodriguez_Fernandez_JPP_2020}.

\section{Scenario optimization} \label{sec6}

Given the essentially quasi-linear nature of the underlying trigger mechanism of the transport barrier observed in the previous sections, scenario optimization is here performed via linear GENE simulations. The parameter space explored consists of the fast ion density, temperature and its logarithmic (temperature) gradient. This analysis aims at identifying the optimal fast ion parameters that maximize their beneficial effect on plasma confinement. All the other plasma parameters are kept fixed to the nominal scenario (Case 1) with the only exception of the electron density, that is adjusted self-consistently to satisfy plasma quasi-neutrality. 

We begin by performing scans over the ratio between the energetic particle and the electron density at $\rho_{tor} = 0.5$. This is the radial location where we observed the strongest linear stabilization of ITG growth rates (see Fig.~\ref{fig:fig2}-~\ref{fig:fig3}) and the center of the transport barrier in the nonlinear simulations (see e.g., Fig.~\ref{fig:fig4}).
\begin{figure*}
\begin{center}
\includegraphics[scale=0.40]{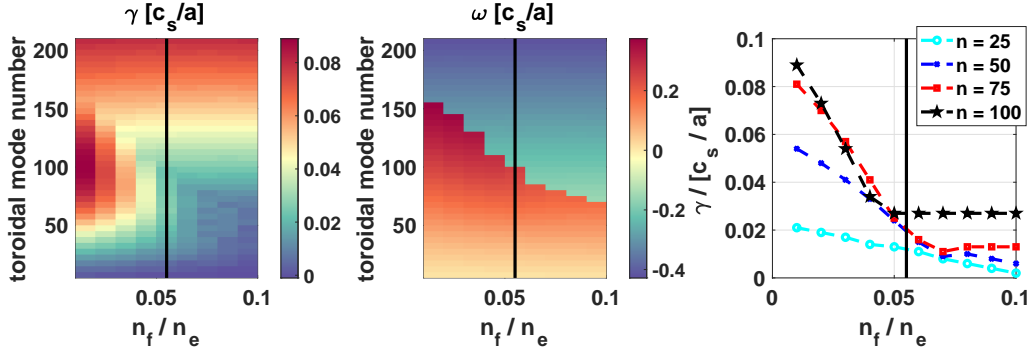}
\par\end{center}
\caption{Contour plots at $\rho_{tor} = 0.5$ of the most unstable (a) linear growth rate and (b) frequency (in units of $c_s / a$ with $c_s = $ the sound speed and $a$ minor radius) for the toroidal mode numbers $n = [1-200]$ at different ratio of the fast ion to electron density $n_f / n_e$. Slices of (a) are shown in (c) for four selected mode numbers. The vertical black line indicates the nominal value of the hydrogen density for the simulations in the previous sections.}
\label{fig:fig10}
\end{figure*}
The growth rates and frequencies  - normalized to the ratio of the sound speed - are shown, respectively, in Fig.~\ref{fig:fig10}a and Fig.~\ref{fig:fig10}b for the toroidal mode number $n = [1 - 200]$. When looking at Fig.~\ref{fig:fig10}, we observe a progressive ITG stabilization as the energetic particle concentration is increased. This is particularly relevant for the range of toroidal mode numbers $n = [20 - 100]$, which are the dominant modes for the nonlinear fluxes. 

The optimal concentration to maximize the beneficial fast ion effect on ITG micro-instabilities for the mode numbers more relevant in the nonlinear simulations is $n_f / n_e = 0.07$. Above this value, ITG is fully suppressed and TEM becomes the dominant instability (see Fig.~\ref{fig:fig10}b), thus making any wave-particle resonant effects negligible. Therefore, an increased fast particle concentration (above $n_f / n_e = 0.07$) does not lead to any further stabilization on the underlying micro-instabilities. This can be seen in Fig.~\ref{fig:fig10}c, where the ITG growth rates dependence with $n_f / n_e $ is illustrated at different toroidal mode numbers. By increasing the fast particle concentration above the nominal one computed by TRANSP (i.e., $n_f / n_e = 0.055$) to this optimal value $n_f / n_e = 0.07$ we observe an additional $50\%$ reduction on the linear ITG growth rates for the toroidal mode numbers $n < 100$.

\begin{figure*}
\begin{center}
\includegraphics[scale=0.40]{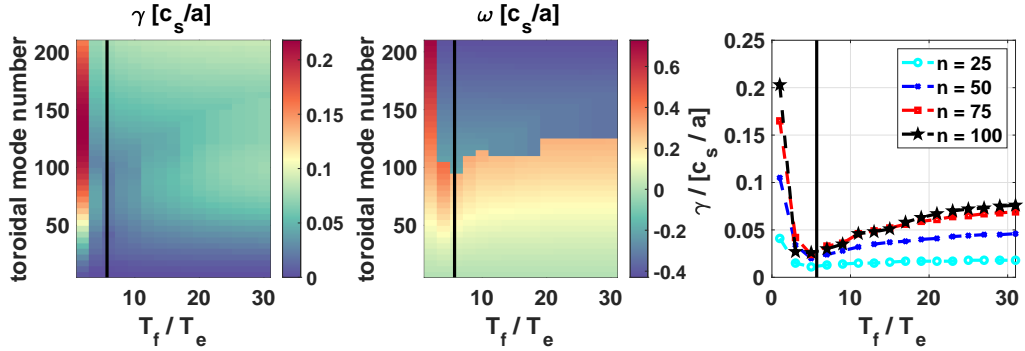}
\par\end{center}
\caption{Contour plots at $\rho_{tor} = 0.5$ of the most unstable (a) linear growth rate and (b) frequency (in units of $c_s / a$) for the toroidal mode numbers $n = [1-200]$ at different ratio of the fast ion to electron temperature $T_f/ T_e$. Slices of (a) are shown in (c) for four selected mode numbers. The vertical black line indicates the nominal value of the hydrogen temperature for the simulations in the previous sections.}
\label{fig:fig11}
\end{figure*}

The impact of the energetic particle temperature on the ITG growth rates is also investigated at the location $\rho_{tor} = 0.5$. The results are illustrated in Fig.~\ref{fig:fig11}, for the toroidal mode number $n = [1 - 200]$. We observe the characteristic "sweet-spot" in the ratio of the energetic particle and electron temperature $T_f / T_e \approx [3 - 6]$ corresponding to the minimum linear growth rates.
If the fast particle temperature is reduced below $T_f / T_e \approx 2$, the resonant layers move in phase-space in the destabilizing fast ion drive region, thus leading to an increase in the ITG growth rates. We notice that the addition of the energetic particle species with $T_f / T_e > 2$ leads to a strong stabilization of ITG micro-instabilities for $n > 100$, where TEM becomes the dominant instability. These results are well consistent with previous studies on different magnetic confinement devices and plasma conditions. Therefore, the hydrogen temperature might be slightly reduced as compared to the reference fast ion temperature profile (see Fig.~\ref{fig:fig1}) reaching $T_h / T_e \sim 9$ at $\rho_{tor} \sim 0.4$. This is a favourable result when combined with the observation from Fig.~\ref{fig:fig10} that the energetic particle density could be increased since an increased fast ion density might be created with the same injected power while still matching the optimal fast ion temperature $T_h / T_e \sim [3 - 6]$. On the other hand, at the base hydrogen concentration of $n_h / n_e = 0.055$, one could reduce ICRF power to maximize Q even further.

\begin{figure*}
\begin{center}
\includegraphics[scale=0.40]{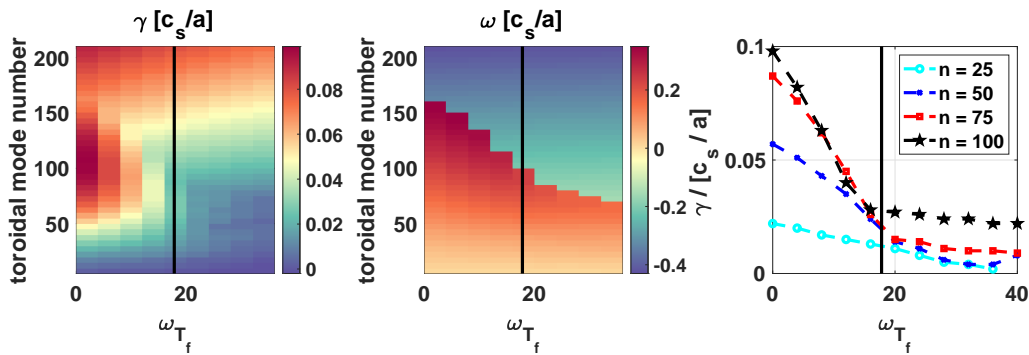}
\par\end{center}
\caption{Contour plots at $\rho_{tor} = 0.5$ of the most unstable (a) linear growth rate and (b) frequency (in units of $c_s / a$) for the toroidal mode numbers $n = [1-200]$ at different fast ion logarithmic temperature gradients $\omega_{T_f}$. Slices of (a) are shown in (c) for four selected mode numbers. The vertical black line indicates the nominal value of the fast ion logarithmic temperature gradient for the simulations in the previous sections.}
\label{fig:fig12}
\end{figure*}

The last fast particle parameter optimized here is the energetic particle temperature gradient $\omega_{T,h}$. It affects the fast particle drive term and hence the direction of the energy exchange between fast particles and ITG micro-instabilities. To identify the optimal value of $\omega_{T,h}$ for this SPARC H-mode scenario, we analyze the ITG growth rate dependence with $\omega_{T,h}$ for the toroidal mode numbers $n = [1 - 200]$. The results are shown in Fig.~\ref{fig:fig12}, for the nominal fast particle concentration and temperature, namely $n_f / n_e = 0.055$ and $T_f/T_e \sim 6$.
Fig.~\ref{fig:fig12}, shows a similar behavior as the one observed in Fig.~\ref{fig:fig10}. More specifically, the growth rates reduce as the fast ion logarithmic temperature gradient increases, which is consistent with the enhancement of the fast ion drive term. The stabilization occurs until fully suppressing ITG, thus making TEM the dominant micro-instability. The optimal value of the fast ion logarithmic temperature gradient maximizing the ITG stabilization for $n < 100$ is $\omega_{T_f} \approx 20$. Above this value, TEM dominates, and fast ion effects on the linear modes become negligible for the toroidal mode numbers considered.

\begin{figure}
\begin{center}
\includegraphics[scale=0.33]{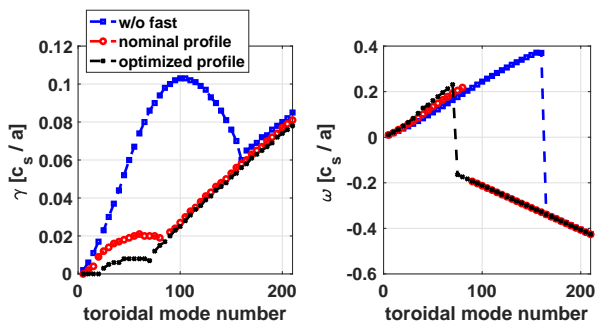}
\par\end{center}
\caption{Linear a) growth rates and b) frequencies (in unit of $c_s / a$) of the simulations retaining (red circles) and neglecting (blue square) the fast particle species modelled with the nominal profiles are compared with the ones obtained with the optimized fast ion parameters (black crosses) at $\rho_{tor} = 0.5$.}
\label{fig:fig13}
\end{figure}

\begin{figure*}
\begin{center}
\includegraphics[scale=0.40]{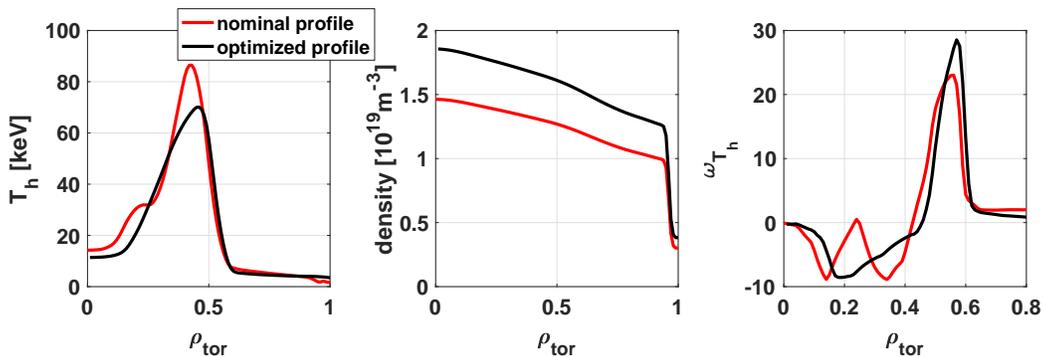}
\par\end{center}
\caption{Energetic particle (a) temperature, (b) density and (c) logarithmic temperature gradient for nominal (see Fig.~\ref{fig:fig1}) and optimized profile.}
\label{fig:fig14}
\end{figure*}

\begin{figure*}
\begin{center}
\includegraphics[scale=0.40]{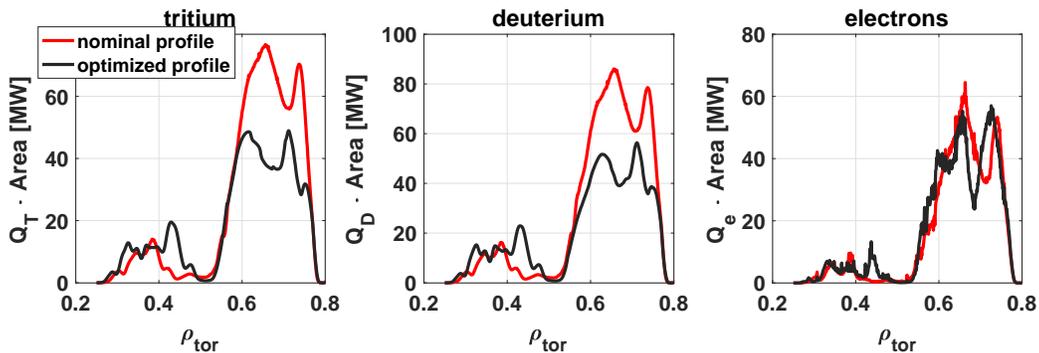}
\par\end{center}
\caption{Radial turbulent heat flux profiles obtained in the global GENE simulations for (a) tritium, (b) deuterium and (c)
electrons averaged over the time domain $t[c_s/a] = [300–370]$ retaining the fast particle species for the nominal (red) and optimized (blue) fast particle pressure profile.}
\label{fig:fig15}
\end{figure*}

The previous scans allowed us to identify the optimal parameters for maximizing the beneficial wave-particle resonance effect on ITGs. In particular, they are $n_f / n_e = 0.07$, $T_f / T_e = [3 - 6]$ and $\omega_{T_f} \sim 20$. By using these parameters at the radial location $\rho_{tor} = 0.5$, an additional $\sim 65\%$ linear stabilization is achieved compared to the nominal plasma profiles (see Fig.~\ref{fig:fig1}). This is shown in Fig.~\ref{fig:fig13}, where the linear growth rates with this optimal setup are compared with the ones obtained with the nominal parameters (see Fig.~\ref{fig:fig3}c).
These results are further corroborated by performing nonlinear global simulations with an ad-hoc fast ion pressure profile satisfying the optimal parameters at $\rho_{tor} = 0.5$ identified above. The corresponding fast ion density, temperature and logarithmic temperature gradient profiles are illustrated in Fig.~\ref{fig:fig14}.
With these optimized, fast ion profiles, a substantial enhancement of the energetic particle turbulence stabilization can be observed in Fig.~\ref{fig:fig15} where the heat flux profiles are compared with the ones obtained with the nominal profiles. More specifically, the fast ion-induced transport barrier has a stronger effect on turbulence, and its beneficial impact on ITGs extends further outside its radial domain. This leads to a reduction of the radially averaged turbulent fluxes by $\sim 40\%$ for deuterium and tritium with respect to Fig.~\ref{fig:fig6}. On the other hand, we observe a mild destabilization by $\sim 15 \%$ for the electron heat flux. This is consistent with the linear destabilization of TEM as ITGs are suppressed by wave-particle resonant effects as shown in Fig.~\ref{fig:fig13}. These results are particularly encouraging for scenario optimization, showing that the fusion gain of this SPARC H-mode scenario could be further improved by carefully optimizing these fast ion effects on the ITGs.

In this regard, it is worth mentioning that the linear optimization performed within this section has been performed at the radial location $\rho_{tor} = 0.5$ (center of the transport barrier). The optimal fast ion parameters are not expected to change if the radial location is changed. This is shown in Ref.~\cite{Di_Siena_PPCF_2022}, where the parameter dependence of the F-ATB was studied at ASDEX Upgrade for different fast ion temperature profiles. In particular, Ref.~\cite{Di_Siena_PPCF_2022} showed that the F-ATB follows the logarithmic fast ion temperature gradient profile. Therefore, if the fast particle temperature profile is shifted along the radial direction - corresponding experimentally to moving the ICRH deposition profile - the transport barrier is expected to move accordingly. This is an interesting dependence of this transport barrier which will be further explored in the near future via gyrokinetic simulations of this SPARC H-mode scenario. More specifically, building the desired fast particle pressure profiles via ICRF heating in the inner core regions would require less auxiliary heating power, given the reduced plasma volume. This might allow us to increase plasma performance with reduced external power, thus possibly further increasing the fusion gain $Q$. Details of the ICRF heating, such as broadening effects and limitations to the achievable gradients might need to be considered as well in future optimizations.

\section{Conclusions} \label{sec7}

This paper investigates the possible role of supra-thermal particles, generated via ICRF heating, on SPARC via radially global and flux-tube (local) GENE simulations. The scenario selected for the numerical analyses is a reduced field and reduced current discharge with hydrogen minority heated with $25$MW of ICRH power. This specific scenario is designed to move the ICRH deposition layer off-axis to $\rho_{tor} \sim 0.5$ and create a significant fraction of supra-thermal particles with the desired properties to maximize ITG turbulence suppression via resonant effects at mid-radius. The plasma profiles are computed with TRANSP-TGLF-EPED, which predicts a fusion gain of $Q \sim 0.9$. By performing linear and nonlinear gyrokinetic simulations, we demonstrate that ITG turbulence can be fully suppressed in the radial region where the energetic particles reach the optimal conditions to maximize the wave-particle resonant stabilization. In particular, a fast ion-induced anomalous transport barrier is observed in the global GENE simulations retaining the fast particle population. In the proximity of the transport barrier radial boundaries, strong zonal flow activity is observed, in agreement with Ref.~\cite{DiSiena_PRL_2021}. While flux-tube simulations are shown to reproduce qualitatively well the turbulent levels outside/inside the transport barrier, they strongly over-predict turbulence at the transport barrier boundaries. In particular, flux-tube simulations fail to correctly describe the zonal flow layers observed in the global simulations. 

Based on these results, we performed a series of global gyrokinetic simulations to assess qualitatively the on-axis peaking of the thermal ion (deuterium and tritium) temperature profiles, thus the increase in the fusion gain, which could be achieved experimentally due to the presence of the transport barrier. More specifically, we designed different temperature profiles for the thermal ions while keeping all the other profiles fixed to the nominal ones computed by TRANSP-TGLF. We show that the simulations retaining the supra-thermal particle species can lead to qualitatively similar turbulent levels as the ones obtained in the simulations without supra-thermal particles (matching the volume integral of the injected sources in the TRANSP-TGLF simulations) despite a substantial peaking of the thermal ion temperature profiles. In particular, the logarithmic temperature gradient increases by more than a factor of two within the transport barrier location, leading to a large improvement of the overall fusion gain of $\sim 80\%$. A quantitative analysis would require gyrokinetic simulations allowing the plasma profiles to self-consistently evolve due to the combined effect of turbulence and external particle and heat sources. These analyses will be performed in the near future, exploiting the recent GENE-Tango code coupling \cite{DiSiena_NF_2022}.

Interestingly, these findings are not significantly affected by electromagnetic fluctuations, possibly related to the large magnetic field and hence low plasma beta.

Moreover, we explored a wide fast ion parameter space via linear gyrokinetic simulations to identify the optimal combination of hydrogen minority density, temperature and logarithmic temperature gradient to enhance the ITG stabilization via wave-particle resonant effects for the scenario at hand. We found that the fast particle stabilization is maximum for $T_h / T_e \sim [3 - 6]$ and increases with their density and logarithmic temperature gradient until TEM becomes the dominant linear instability, i.e., $n_f / n_e \sim 0.07$ and $\omega_{T_f} \sim 20$. A new off-axis scenario was designed with these optimized fast ion parameters, showing an improved linear growth rate stabilization (by $\sim 65\%$ at $\rho_{tor} = 0.5$) and reduced turbulent fluxes (by $\sim 25\%$ computed on the radially-averaged fluxes), possibly leading to even larger fusion performance. The location of the ICRH deposition layer can also be optimized to generate fast particles with the desired properties with the minimum injected power, thus maximizing the fusion gain further. Modifications on its radial location are expected to shift the F-ATB accordingly as shown in Ref.~\cite{Di_Siena_PPCF_2022}.

We note that the volume average electron density, geometry, and input power for this $B_t = 8.5$T and $I_p = 6.1$MA scenario in SPARC were not optimized for performance but rather for maximizing the turbulence stabilization effect. Further work is needed to find the best combination of tokamak actuators to maximize fusion power and fusion gain in reduced-field, reduced-current scenarios that have, generally, less engineering risk. This work, however, provides robust evidence that turbulence stabilization, via carefully tuned energetic particles profiles (e.g. via off-axis heating), can potentially increase performance in the SPARC tokamak, thus opening the possibility of designing new scenarios with enhanced plasma performance via optimization of the energetic particle density and temperature profiles.

\section*{Acknowledgement}

This work has been carried out within the framework of the EUROfusion Consortium, funded by the European Union via the Euratom Research and Training Programme (Grant Agreement No 101052200 — EUROfusion). Views and opinions expressed are however those of the author(s) only and do not necessarily reflect those of the European Union or the European Commission. Neither the European Union nor the European Commission can be held responsible for them. Moreover, P.~Rodriguez-Fernandez and N.~T.~Howard were funded by Commonwealth Fusion Systems via RPP005 and RPP020. Numerical simulations were performed at the Marconi Fusion supercomputers at CINECA, Italy.

\end{document}